# MaaS and GDPR: an overview

Federico Costantini

Department of Law, University of Udine, Via Treppo 18, 33100 UDINE (ITALY)
federico.costantini@uniud.it

**Abstract.** In MaaS, means of transport are virtualized in mobility resources and provided to users using the Internet. From a legal perspective, this model of ITS raises several concerns with regard to data protection. This contribution, after a short description of MaaS and an introduction to the issues of data protection in ITS, explores the impact of GDPR (General Data Protection Regulation) in the European Union, detecting possible threats and remedies and suggesting a plausible approach.

## 1 Introduction: transport, ITS (Intelligent Transport Systems), MaaS (Mobility as a Service)

Transport, as a social phenomenon, is as old as humankind, yet in recent years there have been significant changes due to the impact of "Information Society".[1]

Transport depends on various factors, that can be physical (e.g. orography, infrastructures), social (e.g. economy, culture, customs) or personal (e.g. motivation, preferences) – and duration – they may be persistent (e.g. street map) or temporary (e.g. weather, public events or pollution) – and it can be carried out in different ways depending on the kind of operator e.g. business carrier or public service), object (e.g. goods or people), means (e.g. on foot, bicycle, car, bus, truck, train, boat or airplane) or purpose (e.g. commercial or private).[2]

A transport system becomes "intelligent", according with European Union law, when information is not incidental or external, but a key element. Indeed, Intelligent

---

[1] Floridi, L. (ed.): The Onlife Manifesto. Being Human in a Hyperconnected Era. Springer, Cham (2015).

[2] For an interesting unifying vision in which ICTs are conceived as a sort of mobility, see Kellerman, A.: Mobility or Mobilities: Terrestrial, Virtual and Aerial Categories or Entities? J. of Trans. Geog. 19.4 (2011) pp. 729–737.

Transport Systems (ITS) are defined by the "ITS Directive" (Directive 2010/40/EU)[3] as "systems in which information and communication technologies are applied in the field of road transport, including infrastructure, vehicles and users, and in traffic management and mobility management, as well as for interfaces with other modes of transport".[4] In ITS data are as important as the road, the wheel or the brake, since they shape the "ecosystem" in which transport "systems" move goods and people. In the information exchange among operators, infrastructures and users, we can detect a "proactive" pattern, since agents adapt their behaviour according to feedback received by their environment.

The cutting-edge version of ITS is called MaaS (Mobility-as-a-Service). In it, the demand for transport is fulfilled providing a set of "virtualized" resources. Given the global urbanization trend, such a model provides an alternative, flexible and sustainable metropolitan transportation system, allowing people to plan their multimodal journey and to travel without a personal car. In other terms, passengers can choose any means of transport to reach their destination, using either public service, private operators or shared vehicles. In MaaS, transport in itself becomes a kind of information, so we could argue that data become even more important than wheels.

From a legal perspective, ITS raises several questions concerning the control of information, and specifically about the security of the technological platform where it is managed, the protection of personal data, the accuracy of data concerning roads, traffic and travel, and the transparency and accountability of all processes involved. In Maas such issues become more relevant due to the fact that, in it, "transport" becomes "mobility". Indeed, if ITS can still be seen a "mass" transport system, since it is provided indiscriminately to entire communities, MaaS aspires to offer "personal" mobility, targeting individual needs. To do so, it requires not only a larger amount of data, but also stronger control of information. This is the crucial element that raises new kinds of legal concerns in Maas with regard to data protection.

In the European Union it is not without impact on ITS (and moreover on MaaS, for the reasons above explained), that the data protection legal framework is going to change next year. Indeed, three laws, adopted on 27 April 2016, will come into effect on 25 May 2018: (1) the Regulation 2016/679/EU[5], or General Data Protection Regulation (GDPR), which repeals Directive 95/46/EC;[6] (2) the Directive 2016/680/EU[7] repealing Council Framework Decision 2008/977/JHA,[8] on criminal

---

[3] Directive 2010/40/EU of 7 July 2010 on the framework for the deployment of Intelligent Transport Systems in the field of road transport and for interfaces with other modes of transport, in O.J. 6.8.2010 L 207, pp. 1–13.

[4] Article 4, par. 1, (1) of Directive 2010/40/EU.

[5] Regulation (EU) 2016/679 of the European Parliament and of the Council of 27 April 2016 on the protection of natural persons with regard to the processing of personal data and on the free movement of such data, and repealing Directive 95/46/EC (General Data Protection Regulation), in OJ L 119, 4.5.2016, pp. 1–88.

[6] Directive 95/46/EC of the European Parliament and of the Council of 24 October 1995 on the protection of individuals with regard to the processing of personal data and on the free movement of such data, in OJ L 281, 23.11.1995, pp. 31–50.

[7] Directive (EU) 2016/680 of the European Parliament and of the Council of 27 April 2016 on the protection of natural persons with regard to the processing of personal data by competent

data and records; (3) the Directive 2016/681/EU[9] implementing new rules on Personal Name Record (PNR). Of the three new EU laws mentioned, the crucial one is the first, not only because it reshapes the entire matter of data protection after more than twenty years of prior legislation, but also because a Regulation, unlike a Directive, in the the European Union has a binding effect directly on member States, citizens, institutions and enterprises.[10]

In this contribution I intend to focus on issues concerning data protection in MaaS, introducing an evaluation of the impact of GDPR on such a set of technologies. Since most of questions concerning data protection have been debated originally in ITS, it seems useful to proceed as follows: (1) as a premise, I offer a short description of the MaaS model; (2) I explain how concerns on data protection emerged in ITS; (3) I analyze the legal consequences of MaaS; (4) I set out some final remarks and a path for future research.

## 2 An overview on MaaS model

Mobility-as-a-Service (MaaS) – or also Transportation-as-a-Service (TaaS) – has spread at a very fast pace since the introduction of its basic concepts two years ago by Sampo Hietanen[11] and Sonja Heikkila.[12]

According to a recent article,[13] MaaS is characterized by four components: (1) infrastructure; (2) data providers; (3) transportation operators; (4) trusted mobility

---

authorities for the purposes of the prevention, investigation, detection or prosecution of criminal offences or the execution of criminal penalties, and on the free movement of such data, and repealing Council Framework Decision 2008/977/JHA, in OJ L 119, 4.5.2016, pp. 89–131.

[8] Council Framework Decision 2008/977/JHA of 27 November 2008 on the protection of personal data processed in the framework of police and judicial cooperation in criminal matters, in OJ L 350, 30.12.2008, p. 60–71.

[9] Directive (EU) 2016/681 of the European Parliament and of the Council of 27 April 2016 on the use of passenger name record (PNR) data for the prevention, detection, investigation and prosecution of terrorist offences and serious crime, in OJ L 119, 4.5.2016, pp. 132–149.

[10] Article 288 TFEU (ex Article 249 TEC): "§. 1.- To exercise the Union's competences, the institutions shall adopt regulations, directives, decisions, recommendations and opinions. §. 2.- A regulation shall have general application. It shall be binding in its entirety and directly applicable in all Member States. §. 3.- A directive shall be binding, as to the result to be achieved, upon each Member State to which it is addressed, but shall leave to the national authorities the choice of form and methods […]", consolidated version of the TFEU in OJ C 326, 26.10.2012, pp. 47-201 (pp. 171–172)

[11] See Hietanen, S.: 'Mobility as a Service' – the New Transport Model? Eurotransport 12.2. ITS & Transport Management Supplement (2014), pp. 2–4.

[12] https://en.wikipedia.org/wiki/Transportation_as_a_Service (accessed 1/3/2017). See the app called "Wimp" used in Helsinki since 2016. See UbiGo in Gothenburg, I. C., Karlsson, M., Sochor, J., Strömberg, H.: Developing the 'Service' in Mobility as a Service: Experiences from a Field Trial of an Innovative Travel Brokerage. Transportation Research Procedia 14 (2016) pp. 3265–3273. For a critical overview of MaaS experiences, see Kamargianni, M. et al.: A Critical Review of New Mobility Services for Urban Transport. ibid.

advisors. It is useful to analyze each of them separately, since they all play an important role in the structure of such a model of ITS.

As for the first, in order to allow the deployment of MaaS, there has to be an interconnected physical infrastructure enabling transfers between different transportation services. Moreover, a high level of connectivity is required. Indeed, since users gain access to the system through an app, a widespread use of smartphones is required.

As regards to the second, in MaaS, customers plan their trip selecting the route among different travel options. Of course, this is allowed by the huge amount of data crowdsourced and real-time updated from transport operators, public institutions and others providers.

Concerning the third, public transportation operators remain incumbent players in MaaS, yet their lack of flexibility has driven the growth of private providers offering specific services (e.g. carpooling). Commercial carriers usually do not depend exclusively on MaaS, since they develop their own channels to reach customers – e.g. through an app – and define their own marketing strategy.[14] It is striking, for the analysis in the following paragraph, that fares could be purchased in different ways: by "pay-as-you-go", a monthly subscription[15] or even special packages.

Regarding the fourth, MaaS is intended to be more advanced than a simple journey planner because there is a unique interface between user and the fulfillment of his or her need for "mobility". As an intermediary, MaaS provides information, reservation, assistance and, in the near future, payment. This operator controls data and processes and, by that, is the main figure in the process.

To sum up, the success of MaaS depends not only on a combination of heterogeneous components, but also on a strategic policy of balance between private sector and public interest[16] and, most importantly, on the involvement of people. Users, in fact, have a key element of the system in their pocket. Without them, the system cannot work.

## 3 ITS and Data Protection

Data security and protection have been taken into consideration by European institutions since the very beginning of ITS. Indeed, in the *ITS Action Plan* of 2008,[17] the fifth "action area" pertains to "*Data security and protection, and liability*

---

[13] Goodall, W. et al.: The Rise of Mobility as a Service. Reshaping How Urbanites Get Around. Deloitte Review 20 (2017), pp. 113–129.

[14] The need of a holistic approach to foster the "urban modal shift" is claimed in Batty, P., Palacin, R., González-Gil, A.: Challenges and Opportunities in Developing Urban Modal Shift. Travel Behav. and Soc. 2. 2 (2015), pp. 109–123.

[15] One study suggests the implementation of a unique agency, "one operator with one fleet and one booking system". See Ambrosino, G. et al.: Enabling Intermodal Urban Transport through Complementary Services: From Flexible Mobility Services to the Shared Use Mobility Agency. Research in Transp. Econ. 30, 3 (2016), pp. 1–6.

[16] Goodall et al., 125.

[17] Action plan for the deployment of Intelligent Transport Systems in Europe, COM/2008/0886 final.

*issues*".[18] The EDPS (European Data Protection Supervisor) expressed concern[19] about the proposed "ITS Directive", which afterwards was enacted with safeguard clauses for data protection,[20] data retention[21] and open data[22] contained in Article 10.[23] Although the protection of personal data was included among the priority areas in the *ITS Action Plan*, it does not appear as such in "ITS Directive", where on the contrary we can find an indirect reference to "open data".[24] Despite such an omission,

---

[18] See Action (5.1): "Assess the security and personal data protection aspects related to the handling of data in ITS applications and services and propose measures in full compliance with Community legislation".

[19] See the Opinion of the European Data Protection Supervisor on the Communication from the Commission on an Action Plan for the Deployment of Intelligent Transport Systems in Europe and the accompanying proposal for a Directive of the European Parliament and of the Council laying down the framework for the deployment of Intelligent Transport Systems in the field of road transport and for interfaces with other transport modes (2010/C 47/02), in O.J. C 47/6 of 25.2.2010. Specifically, on data protection in ITS, See Vandezand, N., Janssen, K.: The ITS Directive: More Than a Timeframe with Privacy Concerns and a Means for Access to Public Data for Digital Road Maps? Comp. Law & Sec. Rev. 28.4 (2012), pp. 416–428.

[20] Directive 95/46/EC of the European Parliament and of the Council of 24 October 1995 on the protection of individuals with regard to the processing of personal data and on the free movement of such data, OJ L 281, 23.11.1995, pp. 31–50.

[21] Directive 2002/58/EC of the European Parliament and of the Council of 12 July 2002 concerning the processing of personal data and the protection of privacy in the electronic communications sector (Directive on privacy and electronic communications), in OJ L 201, 31.7.2002, pp. 37-47, amended by Directive 2006/24/EC of the European Parliament and of the Council of 15 March 2006 on the retention of data generated or processed in connection with the provision of publicly available electronic communications services or of public communications networks and amending Directive 2002/58/EC, in OJ L 105, 13.4.2006, pp. 54–63.

[22] Directive 2003/98/EC of the European Parliament and of the Council of 17 November 2003 on the re-use of public sector information, in OJ L 345, 31.12.2003, pp. 90-96, amended by Directive 2013/37/EU of the European Parliament and of the Council of 26 June 2013 amending Directive 2003/98/EC on the re-use of public sector information Text with EEA relevance, OJ L 175, 27.6.2013, pp. 1–8.

[23] Article 10, Rules on privacy, security and re-use of information: "1. Member States shall ensure that the processing of personal data in the context of the operation of ITS applications and services is carried out in accordance with Union rules protecting fundamental rights and freedoms of individuals, in particular Directive 95/46/EC and Directive 2002/58/EC. 2. In particular, Member States shall ensure that personal data are protected against misuse, including unlawful access, alteration or loss. 3. Without prejudice to paragraph 1, in order to ensure privacy, the use of anonymous data shall be encouraged, where appropriate, for the performance of the ITS applications and services. Without prejudice to Directive 95/46/EC personal data shall only be processed insofar as such processing is necessary for the performance of ITS applications and services. 4. With regard to the application of Directive 95/46/EC and in particular where special categories of personal data are involved, Member States shall also ensure that the provisions on consent to the processing of such personal data are respected. 5. Directive 2003/98/EC shall apply".

[24] The "optimal use of roads, traffic and travel data" is qualified as the first priority area "for the development and use of specifications and standards" in Article 2. Article 4 provides the following definitions: "(14) 'road data' means data on road infrastructure characteristics,

pursuant the implementation of the ITS Action Plan, in 2014 the European Commission recommended the adoption of a "template for privacy impact assessments for ITS applications" and of a "Privacy-by-Design" philosophy.[25]

"Privacy by Design" refers to a set of seven principles[26] which are meant to embed respect for data protection into technological devices in order to inhibit license infringement by users. Although this approach is acknowledged as very promising because of its efficacy, it has been criticized for its vagueness and the lack of clear criteria when it comes to practical application.[27]

## 4 GDPR issues in MaaS

Since data are the key factor of MaaS, establishing clear and fair rules for the control of information is crucial. In this paragraph I classify the main emerging issues according to the scheme set out in the previous paragraph.

If it is true that the "ecosystem" of ITS is made not only by physical but also by technological infrastructure, I could argue that in MaaS a "cultural" infrastructure is also required, since to make it possible a huge shift in users' mindset is required: the replacement of the car with the smartphone as a means of transport. The concept of mobility as a whole, in other words, should become part of citizens' lifestyle.

---

including fixed traffic signs or their regulatory safety attributes; (15) 'traffic data' means historic and real-time data on road traffic characteristics; (16) 'travel data' means basic data such as public transport timetables and tariffs, necessary to provide multi-modal travel information before and during the trip to facilitate travel planning, booking and adaptation". The actions to be undertaken are specified in Annex I.

[25] European Commission, Progress Report and review of the ITS action plan, Accompanying the document Implementation of Directive 2010/40/EU of the European Parliament and of the Council of 7 July 2010 on the framework for the deployment of Intelligent Transport System in the field of road transport and for interfaces with other modes of transport, COM(2014) 642 final, SWD(2014) 320 final, of 21.10.2014. See also European Commission, Analysis of Member States reports, Accompanying the document Report from the Commission to the European Parliament and to the Council, Implementation of Directive 2010/40/EU of the European Parliament and of the Council of 7 July 2010 on the framework for the deployment of Intelligent Transport System in the field of road transport and for interfaces with other modes of transport, COM(2014) 642 final, SWD(2014) 319 final, of 21.10.2014. See also the study Directorate-General Mobility and Transport, European Commission ITS Action Plan, FRAMEWORK CONTRACT TREN/G4/FV-2008/475/01, ITS & Personal Data Protection, Final Report, Amsterdam, October 4th, (2012).

[26] The principles are the following: (1) Proactive not reactive, Preventative not remedial (2) Privacy as the default setting; (3) Privacy embedded into design; (4) Full functionality – positive-sum, not zero-sum; (5) End-to-end security – full lifecycle protection (6) Visibility and transparency – keep it open; (7) Respect for user privacy – keep it user-centric, See Cavoukian, A.: Privacy by Design, Take the Challenge. Ontario: Information and Privacy Commissioner of Ontario, (2009).

[27] See for example van Rest, J., Boonstra, D., Everts, M., van Rijn, M., van Paassen, R.: Designing Privacy-by-Design. In: Preneel, B., Ikonomou, D. (eds.) Privacy Technologies and Policy: First Annual Privacy Forum, APF 2012, Limassol, Cyprus, October 10-11, 2012, Revised Selected Papers, Springer, Berlin, Heidelberg (2014), pp. 55-72.

### 4.1 Infrastructure

This is arguably the main problem. The interface between users and MaaS is their smartphone; through it they exchange data with the service provider. In an ideal workflow, after possibly having set their preferences, they send a request and receive in response one or more "mobility solutions", they pay the ticket (if it is not done already) for the one selected. Besides Leaving aside that, matching the GPS coordinates of the smartphone with its location, the system can track users in their route, allowing it to follow his/her movements in real-time and to detect a pattern.

Since the preliminary analysis on ITS,[28] the real possibility that the user could be not only profiled but also "singled out" has raised many concerns, which become more sensitive in MaaS due the increasing number of interconnected databases. For example, it could be possible to find a pattern in a user's movements to and from healthcare facilities, and so correlate travels to certain diseases. Furthermore, a user's destination could be a cult temple or the office of a syndicate, a political party or a civil organization. It is debatable how data concerning such trips could be lawfully treated in MaaS, since in the first cases they would qualify as "data concerning health" by Article 4 §. 1 (15) of GDPR[29] and, in the second example, they would fall into the prohibition of Article 9 §.1 of GDPR.[30] In such cases, and in many others, it seems that the guarantees provided by GDPR, either strictly legal (such as the consent of the data subject (Article 7 GDPR)), or technological (such as "Privacy by Design" (Article 25 GDPR)) are not suitable for completely avoiding the risk that the "controller" or the "processor" could be punished accordingly.

### 4.2 Data providers

In MaaS, data are provided by different sources, which could be involved in the platform (e.g. transport operators), or external (e.g. weather forecast services). Such complexity rises issues not only concerning the "quality of information"[31] processed – e.g., if data are accurate, precise, up-to-date, complete – but also regarding the division of responsibility among providers, "controller" and "processor" whenever incidents of any sort could occur. Of course, according to GDPR "joint controllers" can define respective responsibility by means of an arrangement (Article 26), but such a contract cannot overrule the laws which allocate liability (Article 82).[32] Those criteria are very strict but, likewise, very difficult to apply to MaaS.

---

[28] See the recommendations in Eisses, S. van de Ven, T., Fievée, A. Its Action Plan. Final Report framework Contract Tren/G4/Fv-2008/475/01. ITS & Personal Data Protection. (2012), 35 and 125.

[29] Personal "data related to the physical or mental health of a natural person, including the provision of health care services, which reveal information about his or her health status".

[30] Personal "data revealing racial or ethnic origin, political opinions, religious or philosophical beliefs, or trade union membership, and the processing of genetic data, biometric data for the purpose of uniquely identifying a natural person, data concerning health or data concerning a natural person's sex life or sexual orientation".

[31] Floridi, L., Illari, P.: The Philosophy of Information Quality, Synthese Library. Springer, Berlin, Heidelberg (2014).

### 4.3 Transportation operators

There are no specific relevant issues concerning data protection under such perspective. Indeed, such operators provide for example vehicles or parking lots that are the only material component in MaaS. Of course, complex systems include several kind of service (e.g. shuttle buses or driverless cars), in which further risks could emerge, such as information security, that cannot be considered in this contribution due to lack of space.

### 4.4 Trusted mobility advisors

Third party aggregators are often needed for their ability to match different providers with a unique structured offer with many customers in a coherent marketplace. In MaaS, the aggregator is essentially the owner of the technological platform where different datasets converge, and to which all customers address their demand. Among the many duties of this agent, two are worth mentioning because they are imposed in order to foster information security.

The first is the Data Protection Impact Assessment that has to be performed before starting the "processing". The reason for arguing for the need of such elaborate document is that MaaS can be defined as "a systematic monitoring of a public accessible area on a large scale" (Article 35, §.3 (c) GDPR).

The second is the obligation to notify a personal data breach to the supervisory authority within 72 hours (Article 33 §.1 GDPR) and to the data subject "without undue delay" (Article 34 §.1 GDPR).

## Conclusions

In a wider perspective, in the "Information Society" many differences tend to fade: between selling goods (a personal-owned car) and providing services (a personal car leased or rented), between private (commercial transports, such as taxi, vehicle hire with driver) and public (traditional transport operators, such as bus, metro or tram), among autonomous individual vehicles, collective transport, sharing and parking

---

[32] "§.2.- Any controller involved in processing shall be liable for the damage caused by processing which infringes this Regulation. A processor shall be liable for the damage caused by processing only where it has not complied with obligations of this Regulation specifically directed to processors or where it has acted outside or contrary to lawful instructions of the controller. §.3.- A controller or processor shall be exempt from liability under paragraph 2 if it proves that it is not in any way responsible for the event giving rise to the damage".

services. All of them are "virtualized" in clouds of organized resources, processes and agents.[33]

From a general perspective, in "Information Society" being in control of information is a privileged position: very difficult to achieve but easy to maintain. In ITS data are important, of course, but in MaaS they become the core element, because data represent the "intelligence" of the members of the community, the synthesis of their strategies to overcome transport difficulties avoiding unnecessary costs and trying to take advantage of uncertainty. In MaaS, having the control of such data – generated on a daily basis by the community itself – gives tremendous power, which should be carefully controlled.

The fact that such data have a huge economic value makes it important to establish if and how such information should be transferred or shared with other parties for commercial purposes. They are like natural resources, although they are deeply linked to a specific city or territory, since they are not material.

Considering the remarks on the aspects strictly linked to GDPR and the last observation, I would recommend that all operators draw a specific Code of Conduct concerning Data Protection (Article 40 GDPR) and propose a standard certification in this area (Article 42 GDPR). Indeed, both are needed in order to demonstrate compliance with the obligations concerning GDPR.[34] Furthermore, such a Code of Conduct could be very useful in order to solve or contain privacy issues, to create a framework of fair competition among companies, to encourage new businesses, and to set security measures against cyberattacks and unlawful accesses.

In the near future I intend to investigate this path of research in order to verify how the critical aspects of MaaS could be solved and to see if there are any others.

---

[33] On a wider theoretical perspective, See also Sharma, S.: Evolution of as-a-Service Era in Cloud. https://arxiv.org/abs/1507.00939v1.

[34] Article 24 §.3. "Adherence to approved codes of conduct as referred to in Article 40 or approved certification mechanisms as referred to in Article 42 may be used as an element by which to demonstrate compliance with the obligations of the controller".